\documentclass{epl}

\usepackage[hypertex]{hyperref} 

\usepackage{graphicx}
\usepackage{amsfonts}
\usepackage{amsmath, amsthm, amssymb}

\newcommand{\ve}[1]{\mathbf{#1}}

\newcommand{\A}{{\cal{A}}_\vk}
\newcommand{\Kk}{{\cal{K}}_\vk}
\newcommand{\Kp}{{\cal{K}}_\vp}
\newcommand{\bv}{{\cal{K}}_\vk}
\newcommand{\vk}{\ve{k}} 
\newcommand{\vp}{\ve{p}} 
\def\i{\mathrm{i}} 
\newcommand{\pauli}[1][\alpha\beta]{\boldsymbol{\sigma}_{#1}^{\vphantom{\dagger}}}
\newcommand{\e}[1]{\mathrm{e}^{#1}}

\newcommand{\ie}{\textit{i.e. }}

\newcommand{\eg}{\textit{e.g. }}

\title{Dissipationless spin-current between Heisenberg ferromagnets with spin-orbit coupling}
\shorttitle{Dissipationless spin-current}
\author{J. Linder \and M. S. Gr{\o}nsleth \and A. Sudb{\o}}
\institute{Department of Physics, Norwegian University of
Science and Technology, N-7491 Trondheim, Norway.}


\begin{document}

\maketitle
\begin{abstract}
A system exhibiting multiple simultaneously broken symmetries offers the opportunity to influence physical phenomena such as tunneling currents by means of external control parameters. Time-reversal symmetry and inversion symmetry are both absent in ferromagnetic metals with substantial spin-orbit coupling. We here study transport of spin in a system consisting of two ferromagnets with spin-orbit coupling separated by an insulating tunneling junction. A persistent spin-current across the junction is found, which can be controlled in a well-defined manner by external magnetic and electric fields. The behavior of the spin-current for important geometries and limits is studied. 
\end{abstract}
  
Due to the increasing interest in the field of spintronics in recent years \cite{zutic}, the idea of utilizing the spin degree of freedom in electronic devices has triggered an extensive response in many scientific communities. The spin-Hall effect is arguably the research area which has received most focus in this context, with substantial effort being put into theoretical considerations \cite{dyakonov}
as well as experimental observations \cite{wunderlich}. As the ambition of spintronics is to make use of the spin degree of freedom rather than electrical charge, investigations of mechanisms that offer ways of controlling spin-currents are of great interest. Ferromagnetism and spin-orbit coupling are physical properties of a system that crucially influence the behavior of spins present in that system. For instance, the presence of spin-orbit coupling is highly important when considering ferromagnetic semiconductors \cite{ferrosemi}. Such heterostructures have been proposed as devices for obtaining controllable spin injection and manipulating single electron spins by means of external electrical fields, making them a central topic of semiconductor spintronics \cite{rashba2}.  
In ferromagnetic metals, spin-orbit coupling is ordinarily significantly smaller than for semiconductors due to the bandstructure. However, the presence of a spin-orbit coupling in ferromagnets could lead to new effects in terms of quantum transport.

\par
Studies of tunneling between ferromagnets have uncovered interesting physical effects \cite{nogueira, leeANDlee}. Nogueira \etal~predicted \cite{nogueira} that a dissipationless spin-current should be established across the junction of two Heisenberg ferromagnets, and that the spin-current was maximal in the special case of tunneling between planar ferromagnets. Also, there has been investigations of what kind of impact spin-orbit coupling constitutes on tunneling currents in various contexts, \eg for noncentrosymmetric superconductors \cite{yokoyama}, and two-dimensional electron gases coupled to ferromagnets \cite{wang}. Broken time reversal- and inversion-symmetry are interesting properties of a system with regard to quantum transport of spin and charge, and the exploitment of such asymmetries has given rise to several devices in recent years. For instance, the broken O(3) symmetry exhibited by ferromagnets has a broad range of possible
applications. This has led to spin current induced
magnetization switching \cite{kiselev}, and suggestions have been made
for more exotic devices such as spin-torque transistors
\cite{brataas2} and spin-batteries \cite{brataas3}. It has also led to
investigations into such phenomena as spin-Hall effect in paramagnetic metals \cite{hirsch}, spin-pumping from
ferromagnets into metals, enhanced damping of spins when spins are
pumped from one ferromagnet to another through a metallic sample
\cite{brataas4}, and the mentioned spin Josephson effects in ferromagnet/ferromagnet
tunneling junctions \cite{nogueira}. 

\par
In this Letter, we investigate the spin-current that arises over a tunneling junction separating two ferromagnetic metals with substantial spin-orbit coupling. It is found that the total current consists of three terms; one due to a twist in magnetization across the junction (in agreement with the result of ref. \cite{nogueira}), one term originating from the spin-orbit interactions in the system, and finally an interesting mixed term that stems from an interplay between the ferromagnetism and spin-orbit coupling. After deriving the expression for the spin-current between Heisenberg ferromagnets with substantial spin-orbit coupling, we consider important tunneling geometries and physical limits of our generally valid results. Finally, we make suggestions concerning the detection of the predicted spin-current. Our results indicate how spin transport between systems exhibiting both magnetism and spin-orbit coupling can be controlled by external fields, and should therefore be of considerable interest in terms of spintronics.

\par
\begin{figure}[h!]
\centering
\resizebox{0.75\textwidth}{!}{
\includegraphics{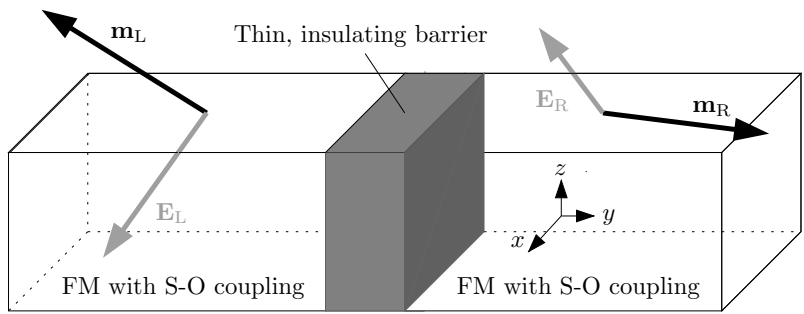}}
\caption{Our model consisting of two ferromagnetic metals with spin-orbit coupling separated by a thin insulating barrier. The magnetization $\mathbf{m}$ and electrical field $\mathbf{E}$ are allowed to point in any direction so that our results are generally valid, while special cases such as planar magnetization etc. are easily obtained by applying the proper limits to the general expressions.}
\label{fig:setup}
\end{figure}

Our system consists of two Heisenberg ferromagnets with substantial
spin-orbit coupling, separated by a thin insulating barrier which is
assumed to be spin-inactive. This is shown in fig. \ref{fig:setup}. A
proper tunneling Hamiltonian for this purpose is $H_\text{T}= \sum_{\vk\vp\sigma}(T_{\vk\vp} c_{\vk\sigma}^\dag d_{\vp\sigma} + \text{h.c.})$,
where $\{c^\dag_{\vk\sigma},c_{\vk\sigma}\}$ and $\{d^\dag_{\vk\sigma},d_{\vk\sigma}\}$ are creation and annihilation operators for an electron with momentum $\vk$ and spin $\sigma$ on the right and left side of the junction, respectively, while $T_{\vk\vp}$ is the spin-independent tunneling matrix element. 
In $\vk$-space, the Hamiltonian describing the ferromagnetism reads
\begin{equation}
H_\text{FM} = \sum_{\vk\sigma}\varepsilon_{\vk} c_{\vk\sigma}^\dag c_{\vk\sigma}- JN\sum_{\vk}\gamma(\vk) \mathbf{S}_\vk\cdot\mathbf{S}_{-\vk}
\end{equation}
in which $\varepsilon_{\vk}$ is the kinetic energy of the electrons,
$J$ is the ferromagnetic coupling constant, $N$ is the number of
particles in the system, $\eta(\vk)$ is a geometrical structure
factor, while $\mathbf{S}_\vk = (1/2) \sum_{\alpha\beta}
c_{\vk\alpha}^\dag\pauli c_{\vk\beta}$ is the spin operator. As we
later adopt the mean-field approximation, $\mathbf{m}=(m_x,m_y,m_z)$
will denote the magnetization of the system.
\par
The spin-orbit interactions are accounted for by a Rashba
Hamiltonian
\begin{equation}
H_\text{S-O} = -\sum_{\vk}
\varphi_\vk^\dag  [\xi(\nabla V\times\vk)\cdot\boldsymbol{\sigma}]\varphi_\vk,
\end{equation}
where $\varphi_\vk = [c_{\vk\uparrow}, c_{\vk\downarrow}]^\text{T}$,
$\mathbf{E}=-\nabla V$ is the electrical field felt by the electrons
and $\boldsymbol{\sigma}=(\sigma_1,\sigma_2,\sigma_3)$ in which
$\sigma_i$ are Pauli matrices, while the parameter $\xi$ is
material-dependent. From now on, the notation
$\xi(\mathbf{E}\times\vk)\equiv\mathbf{B}_\vk =
(B_{x,\vk},B_{y,\vk},B_{z,\vk})$ will be used. In general, the
electromagnetic potential $V$ consists of two parts $V_\text{int}$ and
$V_\text{ext}$ (see \eg ref. \cite{rashba2} for a detailed discussion
of the spin-orbit Hamiltonian). The crystal potential of the material
is represented by $V_\text{int}$, and only gives rise to a spin-orbit
coupling if inversion symmetry is broken in the crystal structure.
Asymmetries such as impurities and local confinements of electrons are
included in $V_\text{ext}$, as well as any external electrical field.
Note that any lack of crystal inversion symmetry results in a
so-called Dresselhaus term in the Hamiltonian, which is present in the
absence of any impurities and confinement potentials. In the
following, we focus on the spin-orbit coupling resulting from
$V_\text{ext}$, thus considering any symmetry-breaking electrical
field that arises from charged impurities or which is applied
externally. Assuming that the crystal structure respects inversion
symmetry, a Dresselhaus term \cite{dressel} is nevertheless easily
included in the Hamiltonian by performing the substitution
$(\mathbf{E}\times\vk)\cdot\boldsymbol{\sigma} \to
[(\mathbf{E}\times\vk) + {\cal{D}}(\vk)]\cdot\boldsymbol{\sigma}$,
where ${\cal{D}}(\vk)$ = $-{\cal{D}}(-\vk)$.

\par
We now proceed to calculate the spin-current that is generated across
the junction as a result of tunneling. Note that in our model, the
magnetization vector and electrical field are allowed to point in
arbitrary directions. In this way, the obtained result for the
spin-current will be generally valid and special cases, \eg thin
films, are easily obtained by taking the appropriate limits in the
final result.

\par
In the mean-field approximation, the Hamiltonian for the right side of
the junction can be written as $H= H_\text{FM}+H_\text{S-O}$, which in
a compact form yields
\begin{equation}\label{eq:H1}
H_\mathrm{R} =H_0 + \sum_{\vk} \varphi_\vk^\dag 
\begin{pmatrix}
\varepsilon_{\vk\uparrow} & -\zeta_\text{R} + \bv \\
-\zeta^\dag_\text{R}  + \bv^\dag & \varepsilon_{\vk\downarrow} 
\end{pmatrix}
\varphi_\vk,
\end{equation}
where $\varepsilon_{\vk\sigma}\equiv\varepsilon_\vk - \sigma
(\zeta_{z,\text{R}} - B_{z,\vk})$ and $H_0$ is an irrelevant
constant. The FM order parameters are $\zeta_\text{R} =
2J\eta(0)(m_{x,\text{R}}-\i m_{y,\text{R}})$ and $\zeta_{z,\text{R}}
= 2J\eta(0)m_{z,\text{R}}$ and $\bv\equiv B_{x,\vk}-\i B_{y,\vk}$.
For convenience, we from now on write $\zeta=|\zeta|\e{\i\phi}$ and
${\cal{K}}_\vk = |\bv|\e{\i\theta_\vk}$. The Hamiltonian for the left
side of the junction is obtained from eq. (\ref{eq:H1}) simply by the doing the
replacements $\vk\to\vp$ and $\text{R}\to\text{L}$.

\par
In order to obtain the expressions for the spin- and charge- tunneling currents, it is necessary to calculate the Green functions. These are given by the matrix ${\cal{G}}_\vk(\i\omega_n) = (-\i\omega_n+\A)^{-1}$, where $\A$ is the matrix in eq. (\ref{eq:H1}). Explicitly, we have that
\begin{equation}
{\cal{G}}_\vk(\i\omega_n)=\begin{pmatrix}
 G_\vk^{\uparrow\uparrow}(\i\omega_n)& F_\vk^{\downarrow\uparrow}(\i\omega_n)\\
F_\vk^{\uparrow\downarrow}(\i\omega_n) & G_\vk^{\downarrow\downarrow}(\i\omega_n)\\
\end{pmatrix}
\end{equation}
Above, $\omega_n = 2(n+1)\pi/\beta, n=0,1,2\ldots$ is the fermionic Matsubara frequency and $\beta$ denotes inverse temperature. Introducing $X_{\vk}(\i\omega_n) = (\varepsilon_{\vk\uparrow}-\i\omega_n)(\varepsilon_{\vk\downarrow}-\i\omega_n) - |\zeta_\mathrm{R}-\Kk|^2$, the normal and anomalous Green's functions are
\begin{align}\label{eq:green}
G_\vk^{\sigma\sigma}(\i\omega_n) = (\varepsilon_{\vk,-\sigma} - \i\omega_n)/X_{\vk}(\i\omega_n),\; F_\vk^{\downarrow\uparrow}(\i \omega_n) = F_\vk^{\uparrow\downarrow,\dag}(\i\omega_n) =  (\zeta_\text{R}-\bv)/X_{\vk}(\i\omega_n).
\end{align} 

Defining a spin-current is not as straight-forward as defining a charge-current \cite{shi}. Specifically, the conventional definition of a spin-current given as spin multiplied with velocity suffers from severe flaws in systems where spin is not a conserved quantity. In this Letter, we define the spin-current across the junction as $\mathbf{I}^\text{S}(t) = \langle \text{d}\mathbf{S}(t)/\text{d}t
\rangle$ where $\mathrm{d}\mathbf{S}/\mathrm{d}t = \i[H_\mathrm{T},\mathbf{S}]$. It is then clear that the concept of a spin-current in this context refers to the rate at which the spin-vector $\mathbf{S}$
on one side of the junction changes \textit{as a result of tunneling
  across the junction}. We choose the right side, denoting the physical parameters with labels $\vk$ and R. In this way, we avoid non-physical interpretations of the spin-current in terms of real spin transport as we only calculate the contribution to $\mathrm{d}\mathbf{S}/\mathrm{d}t$ from the tunneling Hamiltonian \textit{instead} of the entire Hamiltonian $H$. Should we have chosen the latter approach, one would run the risk of obtaining a non-zero spin-current due to \eg local spin-flip processes which are obviously not relevant in terms of real spin transport across the junction. 
\par 
The expression for $\mathbf{I}^\text{S}(t)$ is established by first considering the generalized number operator $N_{\alpha\beta} = \sum_\vk c_{\vk\alpha}^\dag c_{\vk\beta}$. This operator changes with time due to tunneling according to $\dot N_{\alpha\beta} = \i[H_\text{T},N_{\alpha\beta}]$, which in the interaction picture representation becomes $\dot N_{\alpha\beta}(t) = -\i\sum_{\vk\vp}(T_{\vk\vp} c_{\vk\alpha}^\dag d_{\vp\beta}\e{\-\i teV} - \text{h.c.}).$ The voltage drop across the junction is given by the difference in chemical potential on each side, \ie $eV = \mu_\text{R}-\mu_\text{L}$. The Matsubara formalism then dictates that the spin-current across the junction is
\begin{equation}
  \mathbf{I}^\text{S}(t) = \frac{1}{2}\sum_{\alpha\beta}\pauli \langle \dot N_{\alpha\beta} (t) \rangle,
\end{equation}
where the expectation value of the time derivative of the transport operator is calculated in the linear response regime by means of the Kubo formula $\langle \dot N_{\alpha\beta}(t)\rangle = -\i \int^t_{-\infty} \text{d}t' \langle [\dot N_{\alpha\beta}(t),H_\text{T}(t')]\rangle$.

\par
Consider now the $z$-component of the spin-current in particular, which can be written as $I^\text{S}_z = \Im\text{m}\{\Phi(-eV)\}$. The Matsubara function $\Phi(-eV)$ is found by performing analytical continuation $\i p_n \to -eV + \i 0^+$ on $\widetilde{\Phi}(\i p_n)$, where
\begin{align}\label{eq:mat}
\widetilde{\Phi}(\i p_n)=& \frac{1}{\beta}\sum_{\i \omega_m, \vk\vp} \sum_\sigma \sigma \Big(  G_\vk^{\sigma\sigma}(\i\omega_m) G^{\sigma\sigma}_\vp(\i\omega_m -\i p_n) + F^{-\sigma,\sigma}_\vk( \i\omega_m)F^{\sigma,-\sigma}_\vp(\i\omega_m -\i p_n) \Big).
\end{align}
Here, $p_n = 2n\pi/\beta$, $n=0,1,2\ldots$ is the bosonic Matsubara frequency.
Inserting the Green's functions from eq. (\ref{eq:green}) into eq. (\ref{eq:mat}), one finds that a persistent spin-current is established across the tunneling junction. For zero applied voltage, we obtain
\begin{subequations}\label{eq:finalspin}
\begin{align}
I^\text{S}_\text{z} &= \sum_{\vk\vp}  \frac{|T_{\vk\vp}|^2J_{\vk\vp}}{2\gamma_\vk\gamma_\vp}\Big[|\zeta_\text{R}\zeta_\text{L}|\sin\Delta\phi + |\Kk\Kp|\sin\Delta\theta_{\vk\vp}\notag\\
&\hspace{0.95in}-|\Kk\zeta_\text{L}|\sin(\theta_\vk-\phi_\text{L}) - |\Kp\zeta_\text{R}|\sin(\phi_\text{R}-\theta_\vp)\Big], \\
J_{\vk\vp} &= \sum_{\substack{\alpha=\pm\\\beta=\pm}} \alpha\beta\Bigg[\frac{n(\varepsilon_{\vk}+\alpha\gamma_\vk)-n(\varepsilon_\vp+\beta\gamma_\vp)}{(\varepsilon_\vk+\alpha\gamma_\vk)-(\varepsilon_\vp+\beta\gamma_\vp)}\Bigg]. 
\end{align}
\end{subequations}
In eqs. (\ref{eq:finalspin}), $\Delta\theta_{\vk\vp} \equiv \theta_\vk - \theta_\vp$, $\Delta\phi \equiv \phi_\text{R}-\phi_\text{L}$, while $\gamma_\vk^2= (\zeta_{z,\text{R}} - B_{z,\vk})^2 +
|\zeta_\text{R} - \Kk|^2$ and $n(\varepsilon)$ denotes the Fermi
distribution. In the above expressions, we have implicitly associated
the right side R with the momentum label $\vk$ and L with $\vp$ for
more concise notation, such that \eg $B_{z,\vk} \equiv
B_{z,\vk}^\text{R}$. The spin-current described in eq.
(\ref{eq:finalspin}) can be controlled by adjusting the relative
orientation of the magnetization vectors on each side of the junction,
\ie $\Delta\phi$, and also responds to a change in direction of the
applied electric fields. The presence of an external magnetic field
$\mathbf{H}_i$ would control the orientation of the internal
magnetization $\mathbf{m}_i$. Alternatively, one may also use exchange
biasing to an anti-ferromagnet in order to lock the magnetization
direction. Consequently, the spin-current can be manipulated by the
external control parameters $\{\mathbf{H}_i,\mathbf{E}_i\}$ in a
well-defined manner. This observation is highly suggestive in terms of
novel nanotechnological devices.

\par
We stress that eq. (\ref{eq:finalspin}) is \textit{non-zero} in the
general case, since $\gamma_\vk \neq -\gamma_{-\vk}$ and $\theta_{-\vk}
= \theta_{\vk}+\pi$. Moreover, eq. (\ref{eq:finalspin}) is valid for any orientation of
both $\mathbf{m}$ and $\mathbf{E}$ on each side of the junction, and a
number of interesting special cases can now easily be considered
simply by applying the appropriate limits to this general expression.

\par
\begin{figure}[h!]
\centering
\resizebox{0.90\textwidth}{!}{
\includegraphics{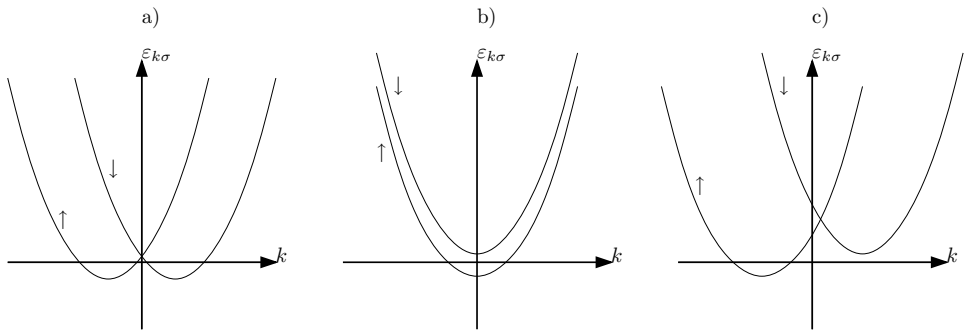}}
\caption{Schematic illustration of the energy-bands for a) a system with spin-orbit coupling, b) a system with ferromagnetic ordering, and c) a system exhibiting both of the aforementioned properties. Since the density of states $N^\sigma(\varepsilon_{k\sigma})$ is proportional to $(\partial\varepsilon_{k\sigma}/\partial k)^{-1}$, we see that a difference between $N^\uparrow(\varepsilon_{k\sigma})$ and $N^\downarrow(\varepsilon_{k\sigma})$ is zero in a), while the density of states differ for the $\uparrow$- and $\downarrow$-populations in b) and c). Thus, a persistent spin-current will only occur for tunneling between systems corresponding to b) and c).}
\label{fig:energy}
\end{figure}

 Consider first the limit where ferromagnetism is absent, such
that the tunneling occurs between two bulk materials with spin-orbit
coupling. Applying $\mathbf{m}\to 0$ to eq. (\ref{eq:finalspin}), it
is readily seen that the spin-current vanishes for any orientation of
the electrical fields. Intuitively, one can understand this by
considering the band structure of the electrons and the corresponding
density of states $N(\varepsilon)$ when only spin-orbit coupling is
present, as shown in fig. \ref{fig:energy}. Since the density of
states is equal for $\uparrow$ and $\downarrow$ spins, one type of
spin is not preferred compared to the other with regard to tunneling,
resulting in a net spin-current of zero. Formally, the vanishing of
the spin-current can be understood by replacing the momentum summation
with integration over energy, \ie $\sum_{\vk\vp} \to \int\int
\mathrm{d}\varepsilon_\text{R}
\mathrm{d}\varepsilon_\text{L}N_\mathrm{R}(\varepsilon_\text{R})N_\mathrm{L}(\varepsilon_\text{L})$.
When $\mathbf{m}\to 0$, eq. (\ref{eq:finalspin}) dictates that 
\begin{align}\label{eq:integral}
  I^\mathrm{S}_z & \sim \sum_{\substack{\alpha=\pm\\\beta=\pm}}
  \alpha\beta \int \int
  \mathrm{d}\varepsilon_{\mathrm{R},\alpha}\mathrm{d}\varepsilon_{\mathrm{L},\beta}
  N^\alpha_\mathrm{R}(\varepsilon_{\mathrm{R},\alpha})N^\beta_\mathrm{L}(\varepsilon_{\mathrm{L},\beta})
  \Bigg[\frac{n(\varepsilon_{\mathrm{R},\alpha})-n(\varepsilon_{\mathrm{L},\beta})}{\varepsilon_{\mathrm{R},\alpha}
    - \varepsilon_{\mathrm{L},\beta}}\Bigg].
\end{align}
Since the density of states for the $\uparrow$- and $\downarrow$-populations are equal in the individual subsystems, \ie $N^\uparrow(\varepsilon)=N^\downarrow(\varepsilon)\equiv N(\varepsilon)$, the integrand of eq. (\ref{eq:integral}) becomes spin-independent such that the summation over $\alpha$ and $\beta$ yields zero. Thus, no spin-current will exist at $eV=0$ over a tunneling barrier separating two systems with spin-orbit coupling alone. In the general case where both ferromagnetism and spin-orbit coupling are present, the density of states at, say, Fermi level are different, leading to a persistent spin-current across the junction due to the difference between $N^\uparrow(\varepsilon)$ and $N^\downarrow(\varepsilon)$.

\par
\begin{figure}[h!]
\centering
\resizebox{0.90\textwidth}{!}{
\includegraphics{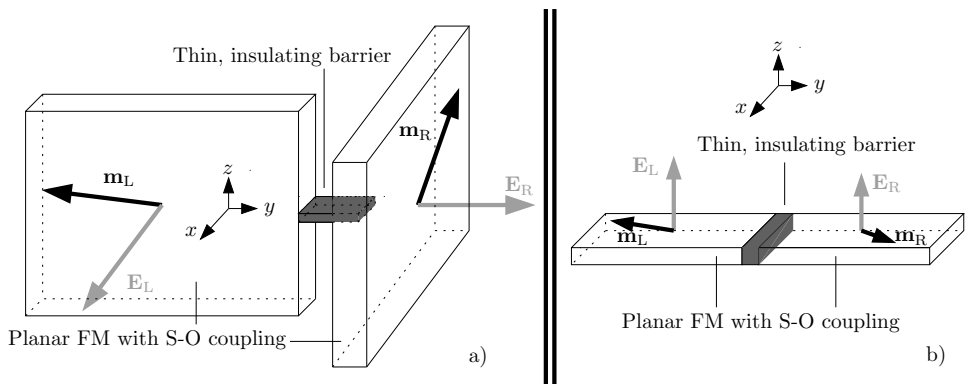}}
\caption{Tunneling between planar ferromagnets in the presence of externally applied electrical fields $\mathbf{E}_\text{L}$ and $\mathbf{E}_\text{R}$ that destroy inversion symmetry and induce a spin-orbit coupling.}
\label{fig:specialcase}
\end{figure}

We now consider a special case where the bulk structures indicated in fig. \ref{fig:setup} are reduced to two thin-film ferromagnets in the presence of electrical fields that are perpendicular to each other, say $\mathbf{E}_\text{L} = (E_\text{L},0,0)$ and $\mathbf{E}_\text{R} = (0,E_\text{R},0)$, as shown in fig. \ref{fig:specialcase}a) and b). In this case, we have chosen an in-plane magnetization for each of the thin-films. Solving specifically for fig. \ref{fig:specialcase}a), it is seen that $\mathbf{m}_\text{L} = (0,m_{y,\text{L}}, m_{z,\text{L}})$ and $\mathbf{m}_\text{R} = (m_{x,\text{R}},0, m_{z,\text{R}})$ . Furthermore, assume that the electrons are restricted from moving in the "thin" dimension, \ie $\vp = (0,p_y,p_z)$ and $\vk = (k_x,0,k_z)$. In this case, eq. (\ref{eq:finalspin}) reduces to the form 
\begin{equation}I^\text{S}_z = I_0\text{sgn}(m_\text{y,L}) + \sum_{\vk\vp} I_{1,\vk\vp}\text{sgn}(p_z)
\end{equation}
where $I_{1,\vk\vp} \neq I_{1,-\vk,-\vp}$; details of the calculations for the setup in fig. \ref{fig:specialcase}a) and b) will be given in a more comprehensive paper \cite{future2}.

\par From these observations, we can draw the following conclusions: whereas the spin-current is zero for the system in Fig. \ref{fig:specialcase}a) if only spin-orbit coupling is considered, it is non-zero when only ferromagnetism is taken into account. However, in the general case where both ferromagnetism and spin-orbit coupling are included, an \textit{additional term} in the spin-current is induced compared to the pure ferromagnetic case. Accordingly, there is an interplay between the magnetic order and the Rashba-interaction that produces a spin-current which is more than just the sum of the individual contributions.

\par
Detection of the induced spin-currents would be challenging,
although recent studies suggest feasible methods of
measuring such quantities. For instance, the authors of ref. \cite{spin2} propose a spin-mechanical device which exploits nanomechanical torque for detection and control of a spin-current. Similarly, a setup coupling the electron spin to the mechanical motion of a nanomechanical system is proposed in \cite{spin1}. The latter method employs the strain-induced spin-orbit interaction of electrons in a narrow gap semiconductor. In ref. \cite{spin5}, it was demonstrated that a steady-state magnetic-moment current, \ie spin-current, will induce a static electric field. This fact may be suggestive in terms of detection \cite{spin3, spin4}, and could be useful to observe the novel effects predicted in this Letter. 

\par
In summary, we have derived an expression for a dissipationless spin-current that arises in the junction between two Heisenberg ferromagnets with spin-orbit coupling. We have shown that the spin-current is driven by terms originating from both the ferromagnetic phase difference, in agreement with the result of ref. \cite{nogueira}, and the presence of spin-orbit coupling itself. In addition, it was found that the simultaneous breaking of time-reversal and inversion symmetry fosters an interplay between ferromagnetism and spin-orbit coupling in the spin-current. Availing oneself of external magnetic and electric fields, our expressions show that the spin-current can be tuned in a well-defined manner. These results are of significance in the field of spintronics in terms of quantum transport, and offer insight into how the spin-current behaves for nanostructures exhibiting both ferromagnetism and spin-orbit coupling.

\acknowledgments

Kjetil B{\o}rkje is acknowledged for very helpful comments. This work was supported by the Norwegian Research Council Grants No. 157798/432 
and No. 158547/431 (NANOMAT), and Grant No. 167498/V30 (STORFORSK).

\end{document}